%% file: nimasic.tex
\documentclass{elsarticle}
\usepackage{graphicx}
\usepackage{epsfig}
\usepackage{amsmath}
\usepackage{amssymb}
\usepackage{epstopdf}
\usepackage{textcomp} 

\usepackage{multirow}

\usepackage{hyperref}

\def\NIMA#1#2#3{{\rm Nucl.~Instr.~and~Meth.} {\bf{A#1}} (#2) #3}

\begin{document}
\begin{frontmatter}
\title{Development of a front end ASIC for Dark Matter directional detection with MIMAC
}
\author[LPSC]{J.~P.~Richer}
\author[LPSC]{G.~Bosson}
\author[LPSC]{O.~Bourrion\corref{cor1}}
\ead{olivier.bourrion@lpsc.in2p3.fr}
\author[LPSC]{C. Grignon}
\author[LPSC]{O. Guillaudin}
\author[LPSC]{F. Mayet}
\author[LPSC]{D. Santos}

\cortext[cor1]{Corresponding author}
\address[LPSC]{Laboratoire de Physique Subatomique et de Cosmologie,\\ 
Universit\'e Joseph Fourier Grenoble 1,\\
  CNRS/IN2P3, Institut Polytechnique de Grenoble,\\
  53, rue des Martyrs, Grenoble, France}

\begin{abstract}

A front end ASIC (BiCMOS-SiGe 0.35~\textmu m) has been developed within the framework of the MIMAC detector project, which aims at directional detection of non-baryonic Dark Matter. 
This search strategy requires 3D reconstruction of low energy (a few keV) tracks with a gaseous \textmu TPC. The development of this front end ASIC is a key point of the project, allowing the 3D track reconstruction. Each ASIC monitors 16 strips of pixels with charge preamplifiers and their time over threshold is provided in real time by current discriminators via two serializing LVDS links working at 320~MHz. The charge is summed over the 16 strips and provided via a  shaper. These specifications have been chosen in order to build an auto triggered electronics.  An acquisition board and the related software were developed in order to validate this methodology on a prototype chamber. The prototype detector presents an anode where $2 \times 96$ strips of pixels are monitored.
\end{abstract} 
\end{frontmatter}

\input{intro}

\input{section2}

\input{conclusion}


\include{biblio}
\end{document}

%% file: intro.tex
\section{Introduction}
Directional detection is a promising non-baryonic Dark Matter search strategy allowing to 
distinguish a genuine WIMP event from  background events. The goal is to show the correlation of the signal with the direction of 
solar motion in the galactic halo \cite{spergel,billard}. This implies a reconstruction of the recoil track down to a few keV.
As the required exposure is decreased by an order of magnitude when going from 2D read-out to 3D read-out \cite{green}, 
3D track reconstruction is needed down to a few keV. 
This can be achieved by using  gaseous micro-TPC detectors, for which  several gases have been suggested :
 $\rm  CF_4,^{3}He+C_4H_{10}, CH_4$ or $\rm C_4H_{10}$ \cite{white,grignon}.
 In particular, the MIMAC (MIcro TPC MAtrix of Chambers) collaboration \cite{grignon,mayet,santos} is planning to build a multi-target detector,
 composed of a matrix of Micromegas gaseous micro-TPC detectors \cite{giomataris,giomataris2}.
\begin{figure}[t] 
\begin{center}
\includegraphics[scale=0.7]{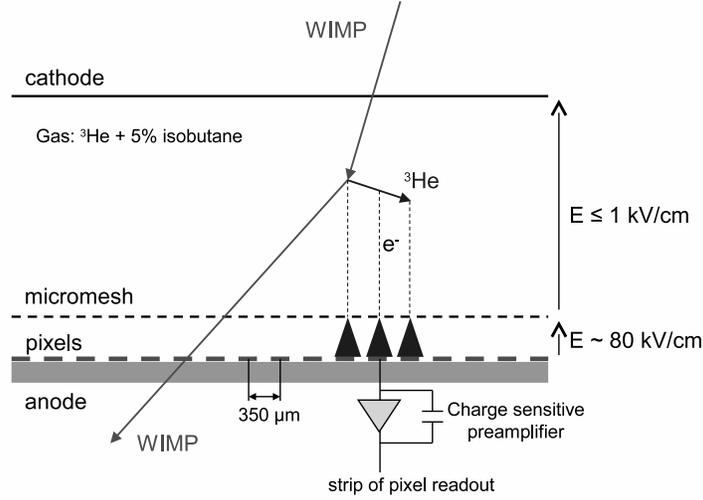}
\caption{Schematic of the MIMAC micro-TPC using a micromegas}
\label{microTPC}
\end{center}
\end{figure}
A pixelized micromegas bulk \cite{giomataris2} is used in order to perform a 3D reconstruction of a few mm track. It 
is segmented in  350\ \textmu m pixels associated to a dedicated ASIC described here. The pixels of each 96 x-rows are linked in jumps of 1 pixel over the 96 y-columns ones forming an interwoven anode of strips of pixels. Indeed, for a given x-row, one out of two pixels is connected to the same strip.  Then, for the next x-row, pixels are connected in a staggered arrangement. Same rule apply for y-columns.
The location of the pixels fired is then obtained by using the coincidence among the x and y strips. This read-out strategy permits to reduce the number of channels for covering a large anode surface as needed for the direct dark matter detection. As shown on figure \ref{microTPC}, the coordinates in the anode plane (x and y coordinates) are reconstructed by collecting primary electrons produced in
the drift region with an electric field ($\leq$1\ kV/cm) and amplified in the avalanche 
region ($\sim$80~kV/cm). The recoil is thus projected on the anode, providing 2D 
information on the track. As stated above, 3D track reconstruction is needed for directional detection of dark 
matter. This is achieved by sampling the anode signal every 25\ ns. Knowing the electron
drift velocity, information on the third coordinate is 
retrieved. For instance, in $\rm ^3He$ at 350\ mbar with a drift field  of 200\ V/cm, the drift velocity is
16\ \textmu m/ns and the recoil track is then decomposed in voxels of typically $350 \times 350 \times 400$\ \textmu $\rm m^3$. A dedicated 3D reconstruction 
algorithm has been developed, allowing to reach an angular resolution better than 
$10^\circ$. More details may be found in \cite{grignon2}.

%% file: section2.tex
\section{Design of the front end ASIC}

\subsection{Background}
The MIMAC detector is planned to be a multi-target detector, i.e. composed of a matrix of chambers filled with several gases. In order to define the ASIC specifications, we focused on $^3$He and $^{19}$F. In the case of $^3$He the following informations have been used:

\begin{itemize}
\item the mean energy to produce a e$^-$ / ion pair is 41 eV;
\item the drift velocity is 16 \textmu m/ns at 350~mbar and a drift field of 200~V/cm;
\item chamber gain is $\sim$5000;
\item minimum signal is $\rm \sim 5000~e^-$ (approximated to a current triangular pulse with a base width of 5~ns and an amplitude of 300~nA );
\item with a drift field of 200~V/cm and a pressure of 350~mbar, the electron cloud radius at one $\sigma$ due to the diffusion is roughly 450~\textmu m\ $\times \sqrt{L(\rm cm)}$ 
where L is the distance from the point of interaction to the anode.
\end{itemize}

Directional detection of dark matter implies the measurement of low energy recoil tracks. 
Typically, for a $^3$He\ \textmu TPC detector, the range of interest for dark matter lies 
below $\sim$10\ keV (50\ keV for $\rm CF_4$), with a sub-keV threshold. Moreover, this \textmu TPC detector 
will be used for neutron metrology up to higher energies (2\ MeV) \cite{amokrane}. Hence, the prototype 
ASIC has been designed and developed with extra functionalities to demonstrate that 
the gaseous chamber associated with its electronics can
measure the energy and track of particles over a wide range of energies.
This will allow the measurement of low energy electrons (used for calibration purposes), low energy nuclear recoils (from WIMP interaction) and even MeV alpha particles. This implies a track measurement with lengths ranging from a few millimetres to a few centimetres. Table \ref{infos} presents ranges and energy losses for various particles.
For the ASIC specification, a dynamic range from 300~eV to 1000~keV was chosen. 
It should be noticed that even for an incoming alpha particle at high pressure the energy channel could saturate and in that case the track length would be enough for discrimination.
\begin{table}[t]
\begin{center}
\begin{tabular}{|c|c|c|}
\hline
\multirow{2}{*}{} & Range                  & \multirow{2}{*}{ Energy} \\
                           & (pressure and gas dependant)&\\

\hline
\multirow{2}{*} {X-source $^{55}$Fe: e$^-$} & 6\ mm (1\ bar) to & \multirow{2}{*}{5.9~keV}\\
											& 18~mm (350~mbar)  & \\

\hline
source $^{57}$Co:  			&  1 mm $\longrightarrow$ 1 cm  & 0.5~keV; 5~keV \\
internal conversion e$^-$   &               (1\ bar)        &  7~keV; 13~keV \\
                          
\hline
ion source:  				&  \multirow{2}{*} {few \textmu m to few mm }  & \multirow{2}{*} {from 0.5 $\longrightarrow$ 50~keV}\\
$^3$He and $^4$He recoils	&											&													\\

\hline
\multirow{2}{*}{neutron capture} & \multirow{2}{*}{$\sim$5~cm (p$^+$) (1 bar)} &  100~keV/cm \\
                 &  &     573~keV + 191~keV\\
\hline
Alpha  & up to chamber size (350~mbar) &  $\sim$5~MeV\\
\hline
\end{tabular}
\caption{Range and energy loss of various particles in the MIMAC detector}
\end{center}
\label{infos} 
\end{table}

The lowest charge that has to be measured by ionization in the dynamic range
chosen is 300\ eV equivalent to 1\ keV recoil at 350\ mbar \cite{SantosQuenching}. It will produce, in the case of $^3\rm He$, 7-8 primary electrons that will drift to
the amplification space  covering a surface determined by the square root of the drift distance in cm. Therefore, the lowest charge to measure will be defined by the gain ($\sim$\ 5000) which has to permit the detection of a primary electron giving 5000\ e$^-$ on one strip of pixels. 
Knowing the approximate shape of the signal, the lowest current to measure is $\sim$300 nA. This narrows the choice of the front end amplification electronics to a charge preamplifier (no current leakage and low noise).

\subsection{Architecture overview}
\begin{figure}[th!]
\begin{center}
\includegraphics[scale=0.5]{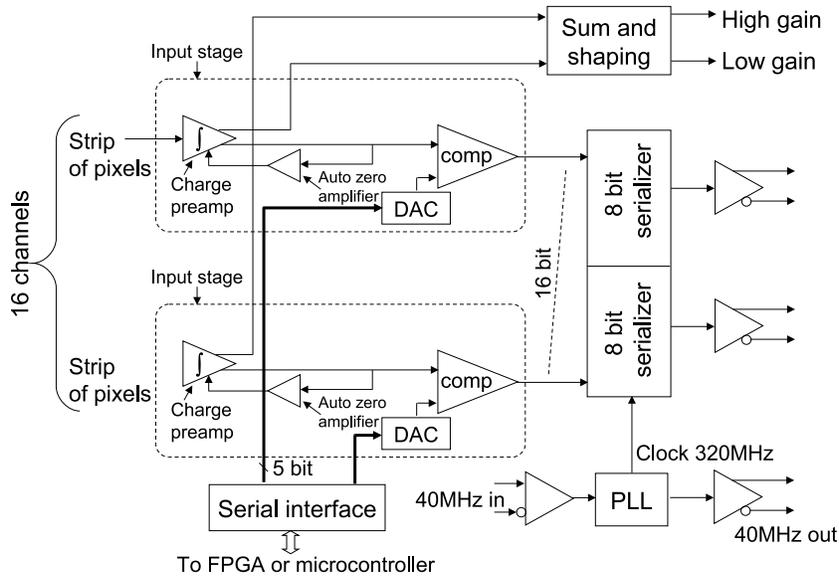}
\caption{Block diagram of the MIMAC ASIC}
\label{asic}
\end{center}
\end{figure}

The front end chip is composed of sixteen channels as presented on figure \ref{asic}. Each channel consists of a charge preamplifier with two outputs. Its voltage output is connected to one of the sixteen inputs of a common summing shaper and its current output feeds a discriminator. The sixteen discriminator outputs are sampled and serialized over two differential lines in order to reduce the system connections. The summing shaper provides the energy released in the detector with two gains. Low gain range is from 300~eV (noise level) to 1~MeV and high gain range is from 300~eV to 80~keV. The position and the inclination of the incident particle track are provided in real time by the digital outputs via the signal duration.
 
\subsection{Input stage}
 
\begin{figure}[th!]
\begin{center}
\includegraphics[scale=0.6,angle=0]{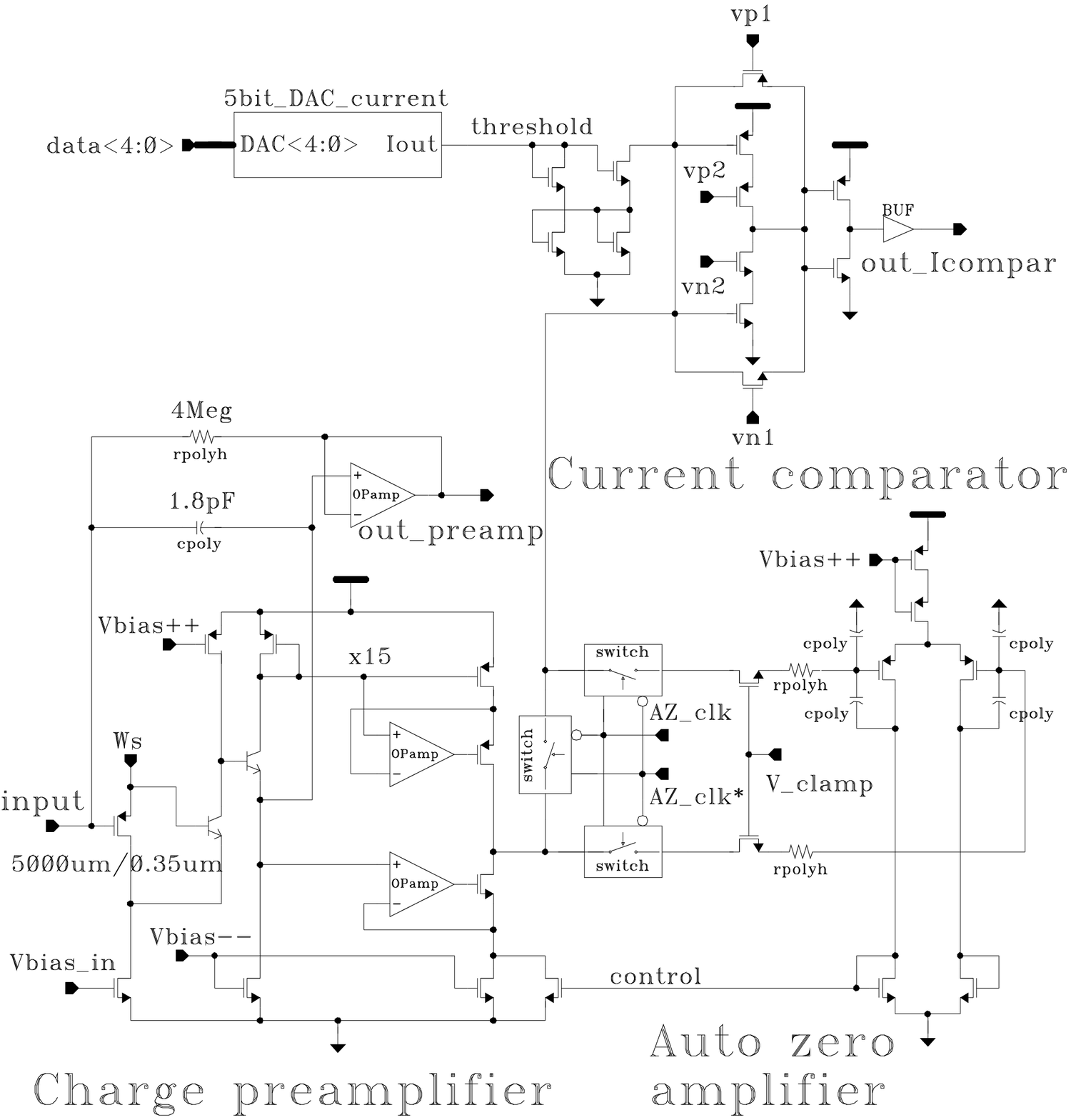}
\caption{Single channel composed of a low noise preamplifier input, an autozero for offset compensation and a comparator}
\label{inputStage}
\end{center}
\end{figure}

In order to maximize the signal to noise ratio, the input stage (see figure \ref{inputStage}) consists of a low noise charge preamplifier based on a folded cascode structure with a large interdigitated PMOS input transistor [5000~\textmu m/0.35 \textmu m].
Due to the fact that the electronics are auto triggered the capacitor is discharged through a resistor with a decay time constant of 8~\textmu s.
To obtain the total energy signal, the voltage output is connected to one of the sixteen shaper inputs to be summed and filtered.

The integration current, flowing through the capacitor, is copied and amplified with a gain of 15 to generate the input signal of the current comparator \cite{linan}. 
Two cascode transistors were added to the CMOS inverter, here used as an amplifier, in order to decrease the influence of the input to output parasitic capacitor. Optimal vn2 and vp2 values were determined by simulation. Also, for improving the comparator commutation speed, the CMOS inverter is kept in linear mode by the use of the two feedback transistors. The other benefit of this architecture is that the injected charges in the transistors grids is minimized, therefore further decreasing the turn on and turn off times. Vn1 and vp1 are adjusted by simulation for barely maintaining the two feedback transistors ``OFF'' in the quiescent state of the comparator (no input current, no threshold applied). When a threshold current (provided by a 5 bit DAC) is applied, one of the two feedback transistors starts conducting allowing the current flow. Then, when an input current higher than the current threshold is applied, the previously conducting transistor is switched off, the CMOS inverter toggles and the other feedback transistor starts conducting. At the output, the comparator delivers a digital signal whose duration is equal to the current duration. This signal is sampled at a rate of 40~MHz. Biasing voltages ``vnx'' and ``vpx'' are internally generated.\\
In order to be able to use this reduced resolution DAC, a trade off has been made and an auto zero amplifier has been added for minimizing the DC offset current in the output branch of the preamplifier (essentially due to transistor mismatches) and thus optimizing the usable DAC range. 
At a slow rate (1 Hz) the current output of the preamplifier is disconnected from the discriminator and connected to one of the differential amplifier inputs. 
The offset current charges or discharges a capacitor and the amplifier compares the voltage capacitor to the DC input voltage of the current comparator. 
The offset correction is applied to the output branch via a current mirror and the ``control'' line. During the auto-zero DC correction, a certain isolation to the input signal has to be implemented in order to provide adequate offset cancellation. For this, series resistors and long channel transistors have been added. The gate voltage ``V\_clamp'', which is accessible outside of the ASIC, provides an adjustment of the filtering time constant.

\subsection{Energy channel}
\begin{figure}[t]
\begin{center}
\includegraphics[scale=0.48,angle=-90]{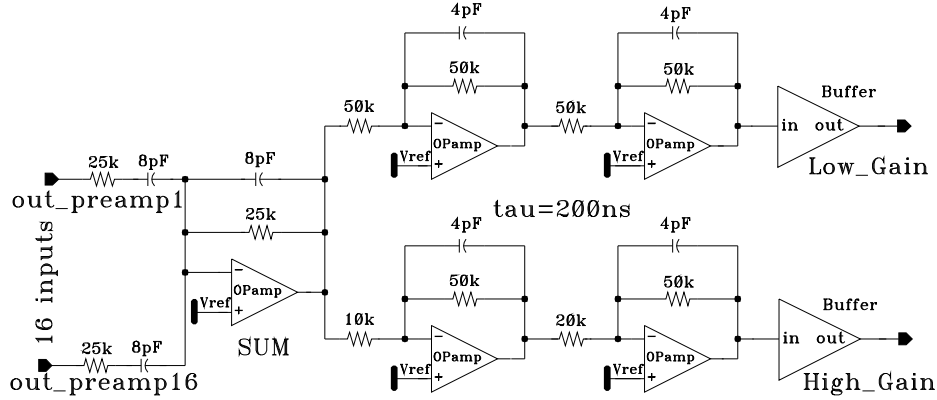}
\caption{Schematic of the shaper having two gains in a ratio of 12.5}
\label{shaper}
\end{center}

\end{figure}

To measure the total energy released in the ionisation channel by the incident particle, the sixteen preamplifier outputs are summed and filtered. 
The signal duration depends on the track inclination, it can last from 10~ns (track of 300~\textmu m in \textsuperscript{3}He at 1 bar) up to 6 \textmu s (track of 15 cm).
 A CR-RC3 structure has been chosen for the shaper with two gains in a ratio of 12.5 and a time constant of 200~ns (figure \ref{shaper}). 
The higher gain increases the precision of the energy measurement in the lower range (up to 80~keV) corresponding to most of the expected experimental results. 
For the time constants the passive components are realized with specific layers available in the SiGe process: high resistivity polysilicon and polysilicon-sinker (bipolar collector) for the capacitor.

\begin{figure}[t]
\begin{center}
\includegraphics[scale=0.35,angle=-90]{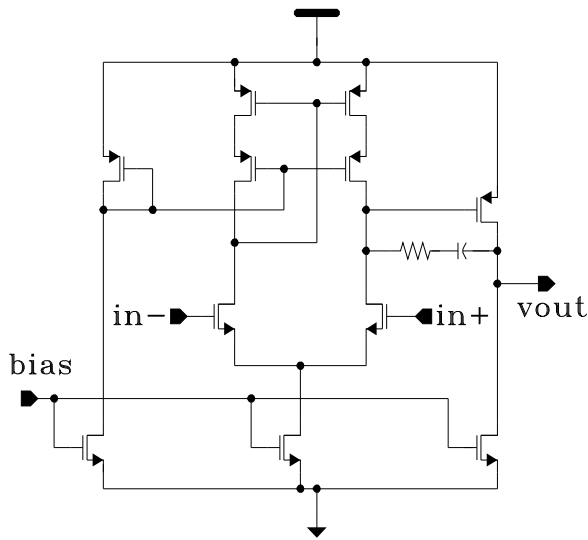}
\caption{Operational amplifier schematic}
\label{SchAmpli}
\end{center}
\end{figure}

\begin{figure}[t]
\begin{center}
\includegraphics[scale=0.50,angle=-90]{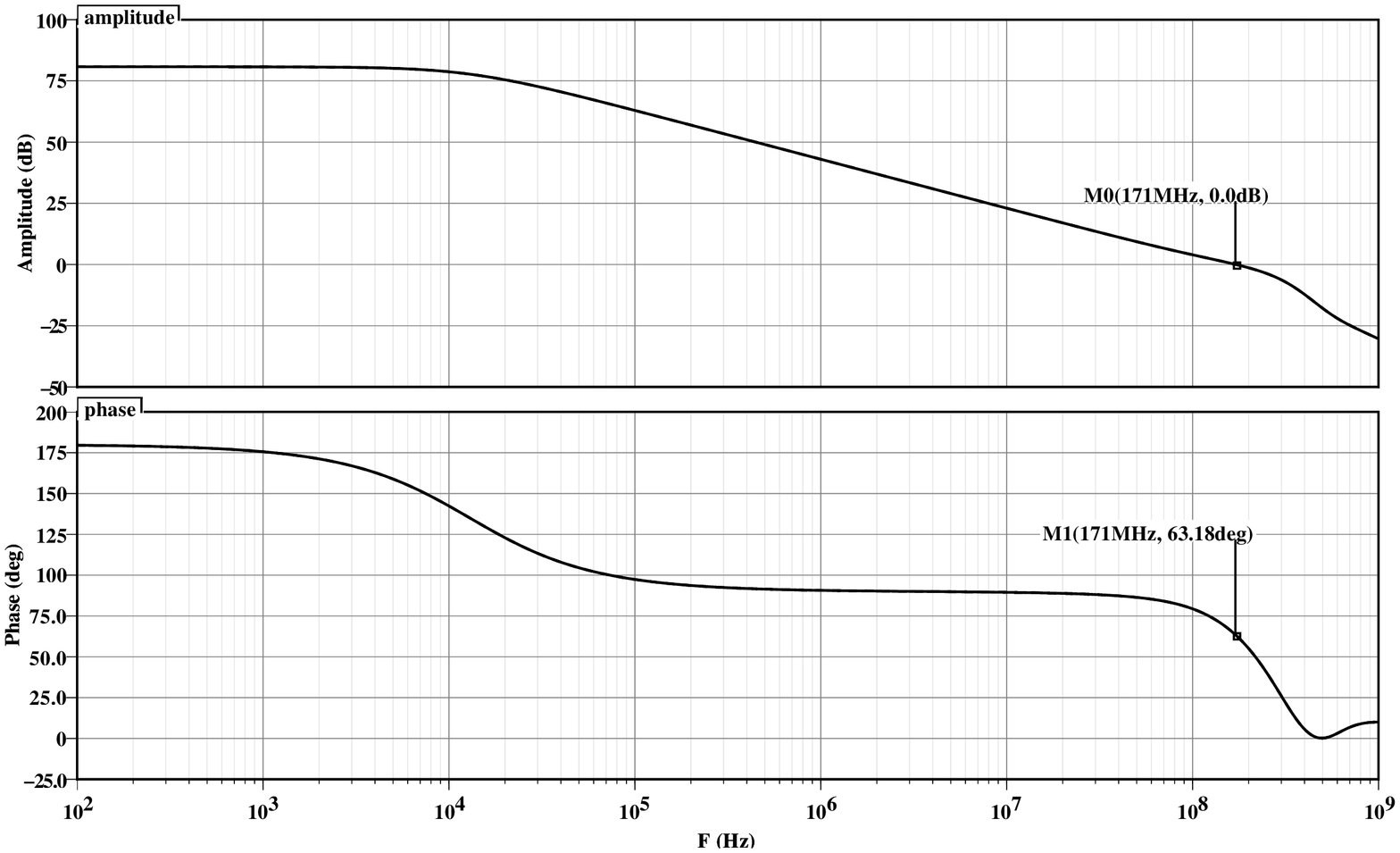}
\caption{Operational amplifier open loop characteristics}
\label{SimuAmpli}
\end{center}
\end{figure}

All the operational amplifiers are based on a similar structure (figure \ref{SchAmpli}) to achieve a gain bandwidth of 170~MHz with a phase margin of 63$^\circ$ and an open loop gain of 80~dB fully compatible with the gains and time constants chosen for the shaper (figure \ref{SimuAmpli}).

\begin{figure}[th!]
\begin{center}
\includegraphics[scale=0.6]{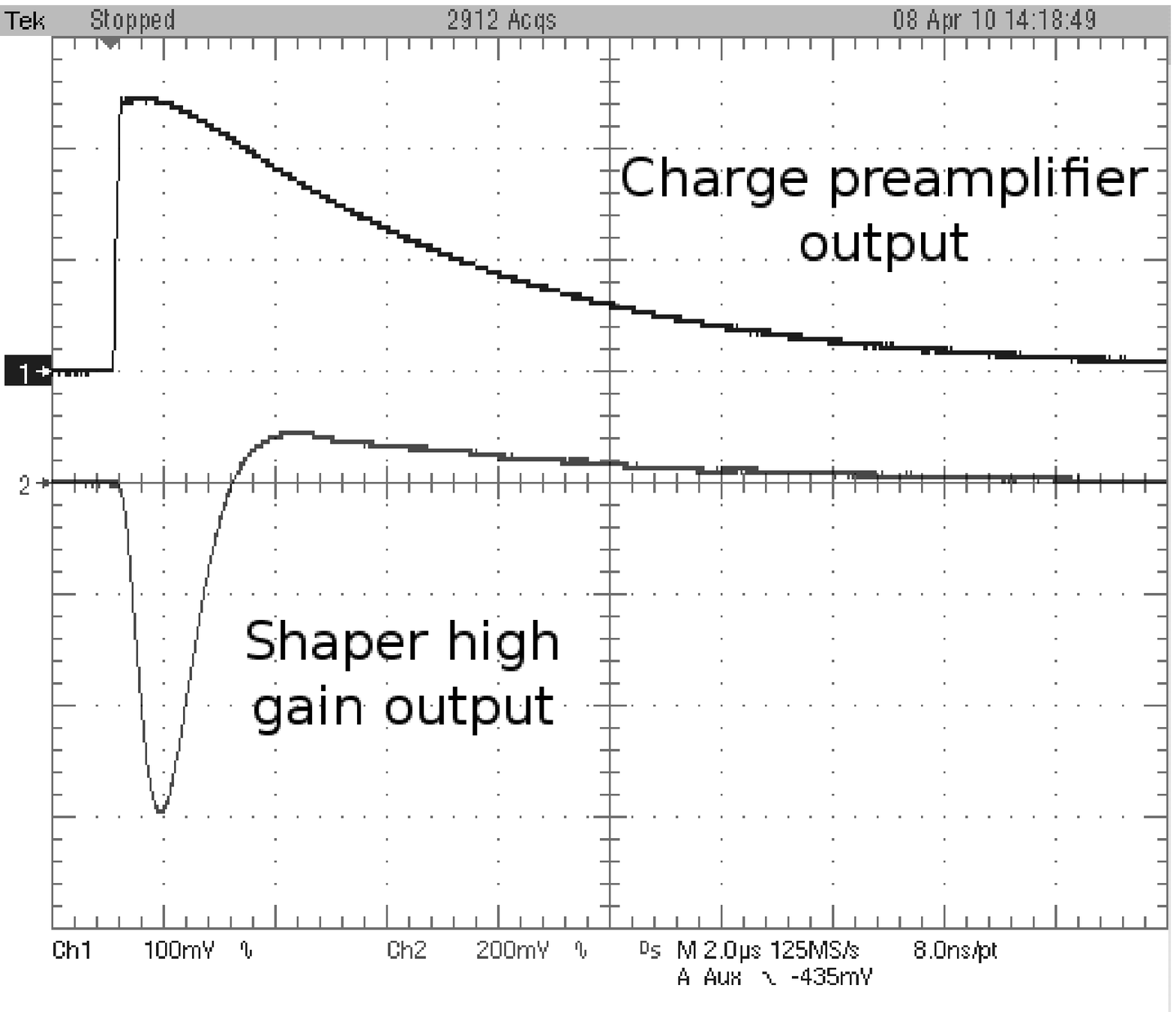}
\caption{Oscillogram of the charge preamplifier and shaper response to a current pulse}
\label{oscillo}
\end{center}
\end{figure}

The oscillogram (see figure \ref{oscillo}) shows the response of the charge preamplifier and the high gain shaper to an input current pulse. 

\begin{figure}[th!]
\begin{center}
\includegraphics[scale=0.52,angle=-90]{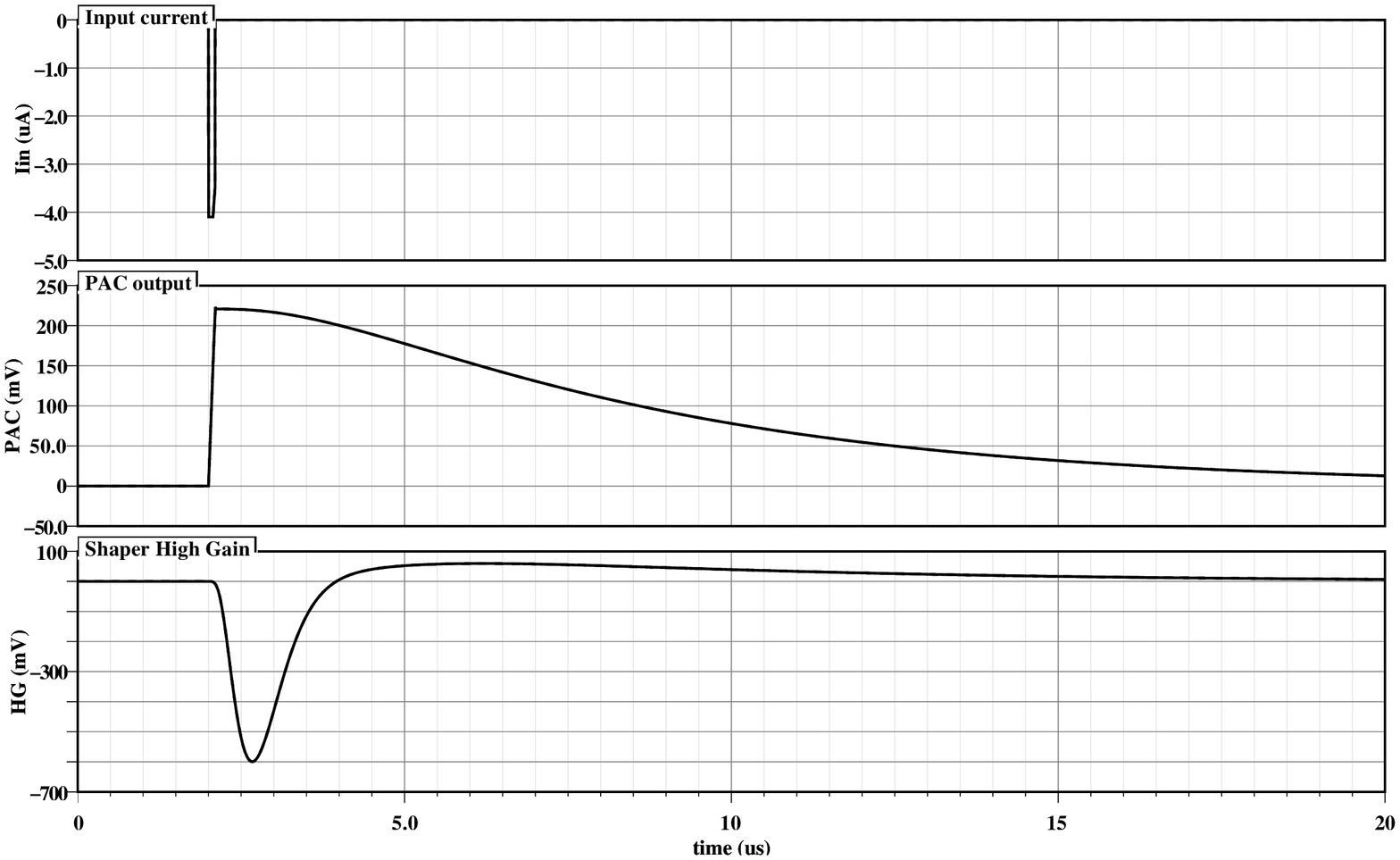}
\caption{Simulation of the charge preamplifier and shaper response to a current pulse}
\label{ShaperSimulation}
\end{center}
\end{figure}

They are in good agreement (with respect to amplitude and peaking time) with the simulation results (figure \ref{ShaperSimulation}). It can be noted that the preamplifier non-exponential decay is due to the parasitic capacitance of the 4~M$\rm \Omega$ feedback resistor (figure \ref{inputStage}), implemented in the high resistivity polysilicon layer, and that it has no negative impact on the shaper output.

\subsection{Serializer}
One major difficulty of this front end ASIC is to get all the comparator outputs out of the chip. This is a concern for several reasons. First, the interconnection has to be kept to a minimum in order to equip large detectors and second, the noise level induced by these outputs has to be kept low. This last reason leads to the choice of LVDS differential output. The problem is that this kind of standard display other difficulties, large number of outputs (two times the number of inputs; one has to remember that the complete MIMAC detector will have around 1000 channels), large current consumption and thus large power dissipation. The remedy to this is to implement a serializer that will work 8 times faster and thus time multiplex data into a single differential pair.

\begin{figure}[th!]
\begin{center}
\includegraphics[scale=0.25]{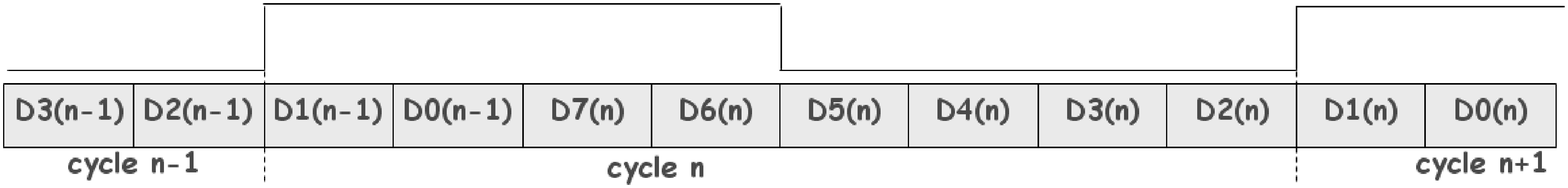}
\caption{Content of the serial frame, each bit is always at the same time offset from the reference clock rising edge}
\label{frame}
\end{center}
\end{figure}

Instead of providing the 8 differential data at a rate of 40~MHz, the data can be transferred on a single pair at a rate of 320~MHz. 
Using this solution, one more signal (differential as well) is required for synchronization, 
it is the 40~MHz reference clock, it avoids to carry an extra signal commutating at 320~MHz.
 When taking a close look at the waveform (see figure \ref {frame}), one can see that each bit is always at the same time offset from the reference clock rising edge. 
This synchronisation signal can be shared by a large number of serializer outputs, as long as each serializer uses the same 320~MHz clock.

\begin{figure}[th!]
\begin{center}
\includegraphics[scale=0.6]{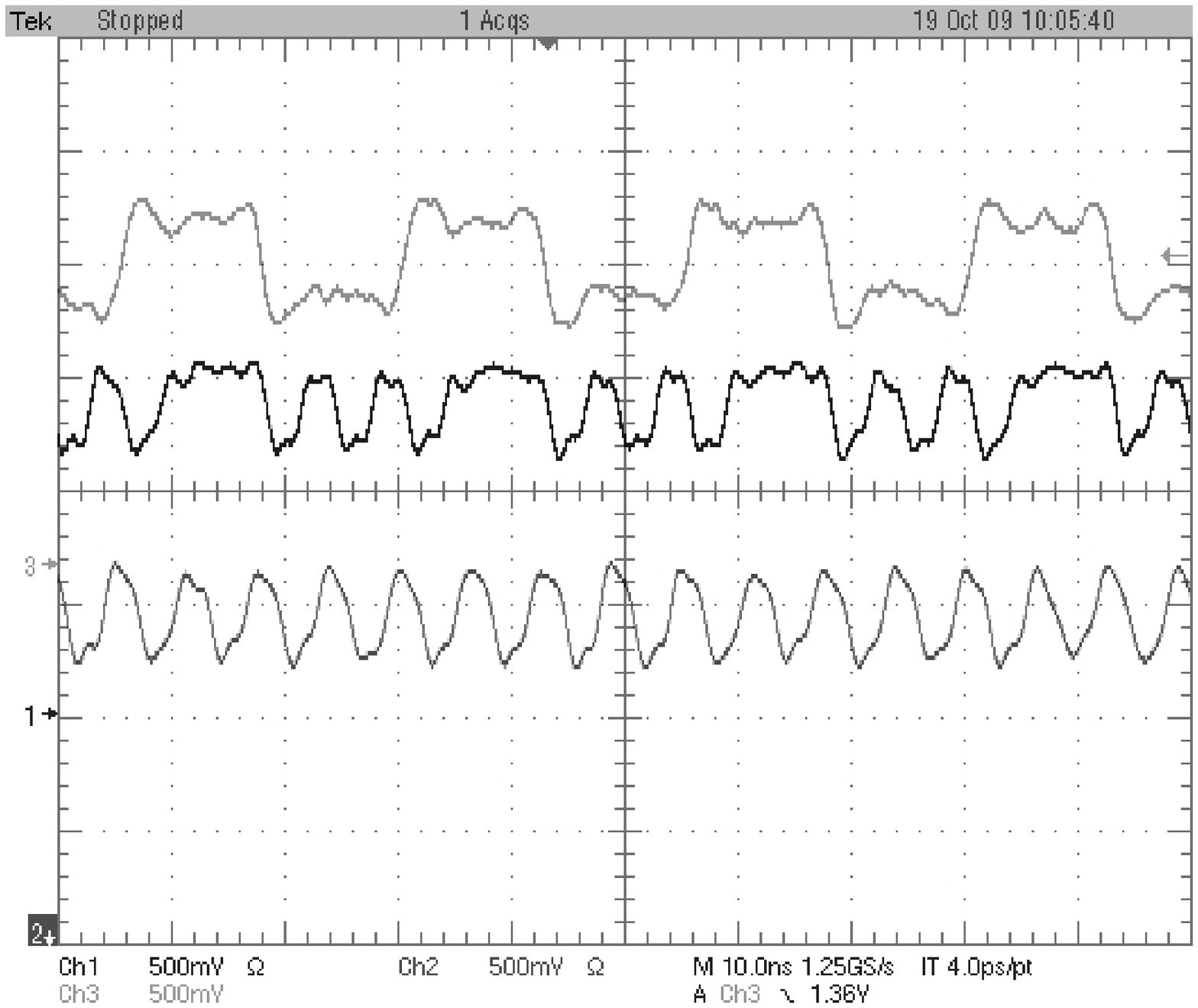}
\caption{Test pattern measurement : top is 40\ MHz clock, middle is serial data 0xEA, bottom is serial data 0xAA}
\label{pattern}
\end{center}
\end{figure}

The 320~MHz clock is generated by a Phase Locked Loop circuit synchronized with the 40~MHz reference clock. 
This ``classical" PLL structure \cite {dzahini} needs only passive external components for the charge pump reference current and filter.
The LVDS transmitter and receiver are based on structures described in \cite {boni}.
For testing purposes, a multiplexer is implemented just before the serializer. That enables the user 
to select either the sending of a known pattern or the comparator output. The oscillogram of the outputs can be seen in figure \ref {pattern}, where a fixed pattern was sent, bottom is serial data 0xAA, middle is the serial data 0xEA, top is the input clock of 40~MHz.

\section{Validation of the front end ASIC with the prototype detector}
In order to validate the front end ASIC design and concept, a $3.36\times 3.36\ \rm cm^2$ prototype anode featuring $2 \times 96$ strips of pixels was coupled to an acquisition board.  It was equipped with $2 \times 6$ ASICs and FPGAs for triggering and hardware level processing. Each strip composed of 96 pixels was connected to one ASIC input channel. The digital outputs were connected directly to the FPGA and the shaper outputs were used to feed peak detectors coupled with ADCs. The FPGA uses the de-serialized front end digital output information for local triggering (an OR of the 16 discriminator outputs). There are as many local triggers and pre processors as ASICs. For each trigger, the FPGA records and decodes each ASIC track informations (pre-processing stage) and retrieves the shaper amplitude from the relevant ADC.\\
In the last stage of the hardware processing, the track and energy informations issued by the ASIC monitoring the involved pixel strips are merged and transferred to the reconstruction and on-line display software.

%% file: conclusion.tex
\section{Conclusion}
\begin{figure}[t]
\begin{center}
\includegraphics[scale=0.5,angle=90]{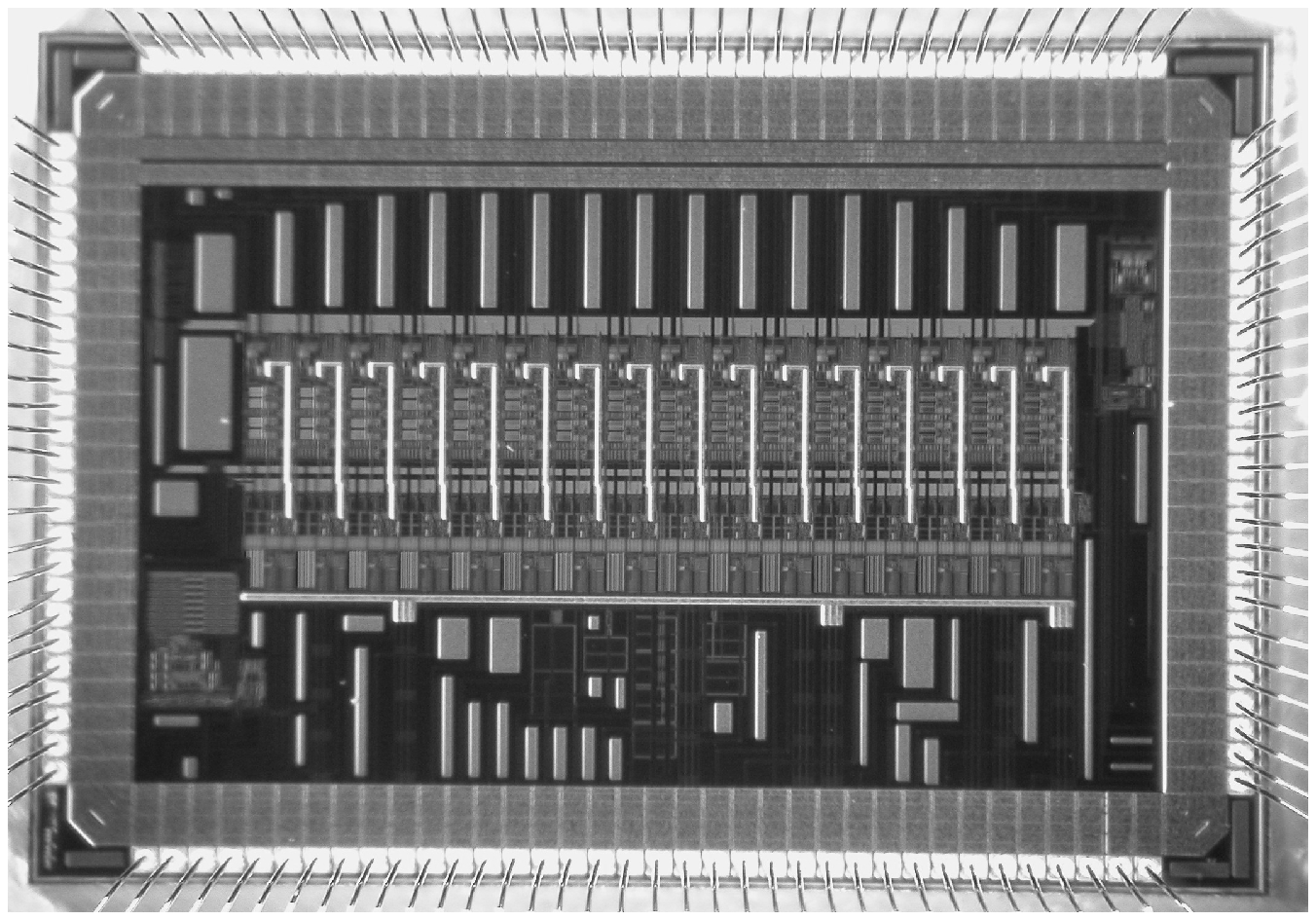}
\caption{Picture of the ASIC (BiCMOS-SiGe 0.35~\textmu m - Area $\sim$15 mm$^2$)}
\label{asicpicture}
\end{center}
\end{figure}

\begin{figure}[t]
\begin{center}
\includegraphics[scale=0.45]{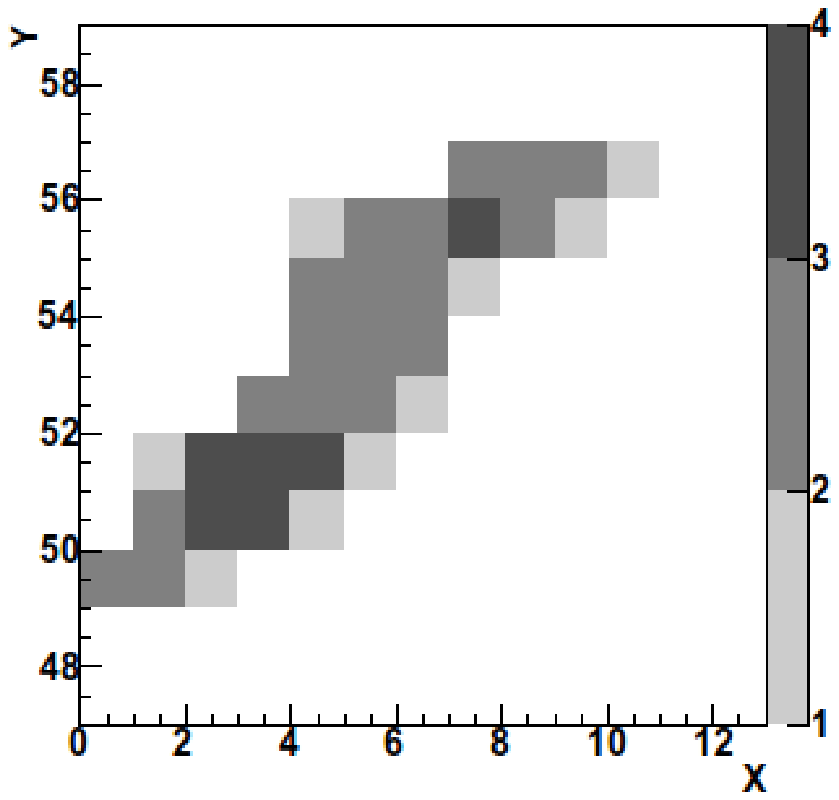}
\includegraphics[scale=0.45]{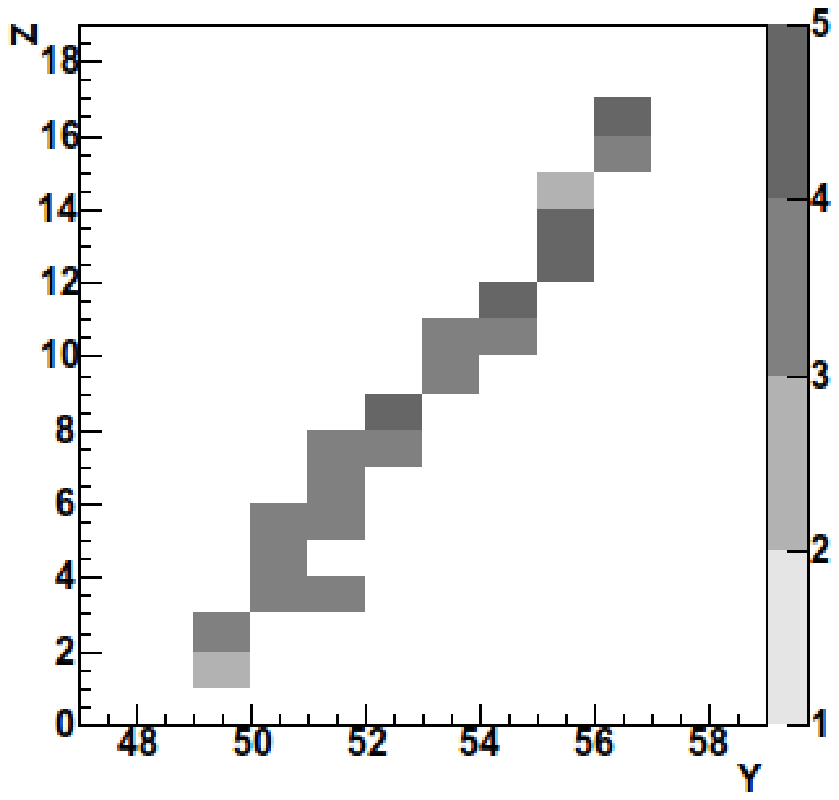}
\includegraphics[scale=0.45]{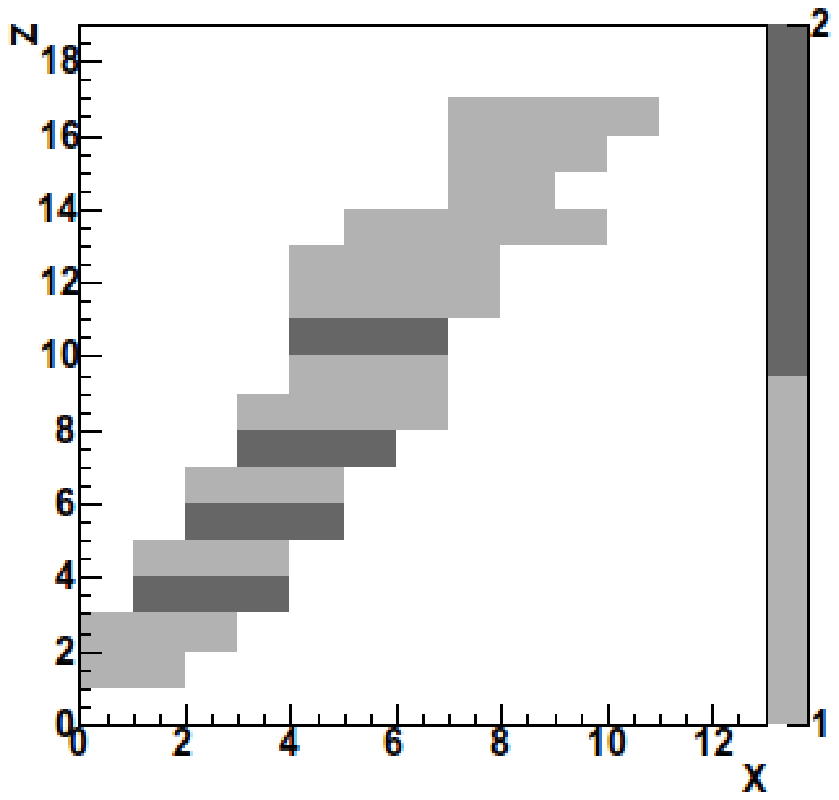}
\caption{3D reconstructed track of a 5.9 keV electron in 95\%~$\rm ^4He+C_4H_{10}$ at a pressure of 350 mbar and a drift field of 200 V/cm.
The left panel presents the XY view of the track: it is a projection of the 17 images (taken every 25 ns) on the anode plane.  Middle and right panel represent the YZ and XZ views of the 3D reconstructed track. Z coordinate is obtained using the number of images and the electron drift velocity.
The grey scale represents the number of time a 2D pixel of the plane, defined by the coincidence of x and y strips, is fired.} 

\label{tracks}
\end{center}
\end{figure}

Twelve of these prototype chips (figure \ref{asicpicture}) have been mounted on a prototype \textmu TPC composed of $2 \times 96$ strips of pixels covering $3.36\times 3.36\ \rm cm^2=11.3\ cm^2$. The detector has been used in various experimental configurations. The data acquisition electronics board and software will be presented in a forthcoming paper \cite{bourrion}. A dedicated 3D reconstruction algorithm is needed to take full advantage of the potential of the detector \cite{grignon2}.
As an example,  a 3D track reconstruction obtained with a MIMAC prototype detector is presented in figure \ref{tracks}. It has been obtained in 95\%~$\rm ^4He+C_4H_{10}$ at a pressure of 350~mbar and with a drift field of 200~V/cm. Figure \ref{tracks} presents this track projected on the X,Y plane (the anode one) on the left panel. The middle (resp. right) panel presents the track measurement  in the  Y,Z (and X,Z) plane. This example outlines the possibility to obtain 3D track measurement down to keV energy particles with a gaseous \textmu TPC equipped with the front end ASIC presented in this paper. 

It has been shown by simulations and experimental results, that the MIMAC ASIC offers great possibilities for 3D track reconstruction and ultimately for 
directional dark matter detection. \\
A 64 channels version of this ASIC is under test, it includes some minor improvements such as:
\begin{itemize}
\item Increased shaper time constant in order to reduce the energy measurement difference between short and long duration tracks
\item An increase of the maximum serial rate in order to reach the 50 MHz sampling rate
\item Removal of the synchronisation clock (known patterns will be used for synchronisation)
\item The addition of a stretcher after each discriminator in order to cope with short duration tracks ($<$ 20 ns)
\end{itemize}

\section*{Acknowledgement}
C.G and the MIMAC collaboration acknowledge the ANR-07-BLAN-0255-03 funding.